\documentclass[journal=jacsat,manuscript=article]{achemso}
\usepackage[version=3]{mhchem} 
\usepackage[dvipsnames]{xcolor}
\pdfoutput=1

\usepackage[normalem]{ulem} 
\definecolor{blue}{rgb}{0,0,1}
\definecolor{cyan}{rgb}{0.2,0.7,1}

\author{Svetlana Melnik (Ponomarenko)\textsuperscript{1}}
\author{Alexander Ryzhov\textsuperscript{1}}
\affiliation{Center for Energy Science and Technology, Skolkovo Institute of Science and Technology, 121205 Moscow, Russia}
\author{Alexei Kiselev}
\affiliation{Institute of Meteorology and Climate Research, Karlsruhe Institute of Technology, 76021 Karlsruhe, Germany}
\author{Aleksandra Radenovic}
\affiliation{Institute of Bioengineering, École Polytechnique Fédérale de Lausanne (EPFL), CH-1015 Lausanne, Switzerland}
\author{Tanja Weil}
\affiliation{Max Planck Institute for Polymer Research, Ackermannweg 10, 55128 Mainz, Germany}
\author{Keith J. Stevenson}
\affiliation{Center for Energy Science and Technology, Skolkovo Institute of Science and Technology, 121205 Moscow, Russia}
\author{Vasily G. Artemov}
\email{vasily.artemov@epfl.ch}
\affiliation{Institute of Bioengineering, École Polytechnique Fédérale de Lausanne (EPFL), CH-1015 Lausanne, Switzerland}

\title{Confinement-controlled Water Engenders High Energy Density Electrochemical-double-layer Capacitance}

\begin{document}


\begin{abstract}

The renewable energy sector critically needs low-cost and environmentally neutral energy storage solutions throughout the entire device life cycle. However, the limited performance of standard water-based electrochemical systems prevents their use in specific applications. Meanwhile, recent fundamental studies revealed dielectric anomalies of water near solid-liquid interfaces of carbon-based nanomaterials. In contrast to the bulk water properties, these anomalies of water under nano-confinement and in the presence of electric fields have not yet been understood and used. Here, we experimentally study the ability of the interfacial water layer to engender and store charge in electrochemical double-layer capacitance. We demonstrate the prototype of a ‘water only’ membrane-electrode assembly. The prototype exhibits characteristics with a perspective of competing with existing batteries and supercapacitors without using electrolytes as ionic carriers. The results provide the impetus for developing high-energy-density electrochemical double-layer capacitors and open up other avenues for ecologically-neutral batteries, fuel cells, and nanofluidic devices.

\end{abstract}

\begin{figure}
\includegraphics{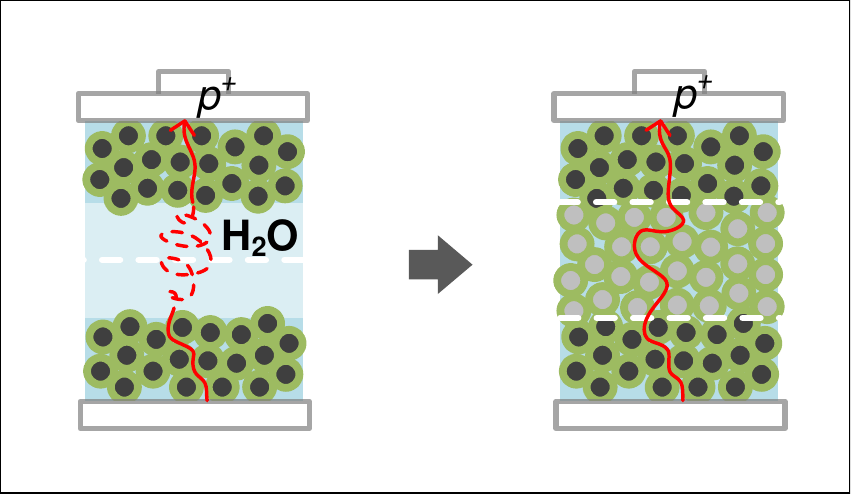}
\centering
\end{figure}

\footnotetext[1]{These authors contributed equally}

\section{Introduction}

State-of-the-art renewable energy networks strongly depend on the availability of electrochemical energy systems, such as batteries, fuel cells, or supercapacitors~\cite{Arbabzadeh2019}. These devices are either direct sources of electricity or temporary reservoirs of electrical energy, though both share a similar design. They consist of a cathode, an anode, and an electrolyte with the difference that supercapacitors store ions at electrified interfaces with no charge transfer across the interfaces (non-faradaic electrochemical double-layer capacitance), while batteries and fuel cells convert and store charge via redox transformations (faradaic ion-coupled, electron transfer process) of chemical species via electrode reactions~\cite{Liu2018, Gogotsi2019, Jiao2021}. An advantage of electrochemical systems compared to conventional fuel-combustion machines is their scalability, allowing demand-dependent dimensions. However, the majority of superior materials for electrochemical energy storage and the way of fabrication are not sustainable, and their exponentially growing turnout poses serious recycling and environmental concerns all over the world~\cite{Larcher2015, Armand2008, Yoo2014, Dominkovich2018}. These problems can be resolved by using environmentally-neutral materials with reasonable energy storage capacitance.

Energy storage solutions based on aqueous systems offer an attractive alternative owing to their low cost, inherent environmental  neutrality, relatively high power density, and safety~\cite{Posada2017, Blomquist2017, Chao2020}. They are experiencing a come-back to keep up with the accelerated transition to renewable sources of electricity~\cite{Hargadon2012, safaei2015, Davidson2019}, however, a relatively low energy density and hardly-predictable long-term operation stability of existing water-based electrochemical systems based on proton exchange membranes limit their range of applications, prevent a widespread distribution, and give way to less sustainable metal-ionic solutions~\cite{najib2019}. The main problem of water-based systems is an insufficient understanding of the dynamic structure of aqueous interfaces under various mechanical, thermo-, and electro-dynamic conditions~\cite{Gonella2021, Fleischmann2022}. This problem limits the optimization of electrodes and separators of ionic species (membranes), as well as tangle their modeling~\cite{pugach2018, mahankali2019, Li2021}. 

Here, we used state-of-the-art knowledge of the structure and dynamics of confined water to construct a carbon-based porous electrochemical double-layer capacitor. We tested the limits of the charge storage capacity of the device at different pore sizes in the electrodes' separator down to a few nanometers. We found an increased H$^+$ and OH$^-$ ions mobility in water confined in 3 nm pores, allowing for the transfer of a larger amount of charge to the electrode interface compared to the bulk water. We show that connected-via-pure-water-filled-nanochannels carbon-water interfaces on the electrodes allow storing a considerable amount of hydronium (H$_3$O$^+$) and hydroxyl (OH$^-$) ions (hereinafter ‘excess protons’), even in absence of electrolyte. We show that the charge storage capacity of our water-only device is mainly due to the size effect and anticipate the same effects in other materials with the same pore sizes. Our research paves the way for inherently safe, low-cost, and environmentally friendly energy storage based on hydrogen ions.

\section{Main}

\subsection{Confinement-induced distortions of water structure}

Previous studies revealed structural and dynamic anomalies of water near solid-liquid interfaces and showed that the properties of nanometer-thick interfacial layer differ from that of bulk water~\cite{Israelachvili1983, Lee1984, Toney1994, Holt2006, Cicero2008, Tocci2014, Velasco2014, Feng2016, Fumagalli2018, Li2020, Artemov2020a}. When water is placed between surfaces a few nanometers apart, the interfacial water layer becomes isolated from bulk water or confined. The observed anomalies of nano-confined water include changed mechanical, electrodynamic, and thermodynamic properties, which have been revealed experimentally and studied theoretically~\cite{Hu2014, Feng2016, Bjorneholm2016, Secchi2016, Lozada-Hidalgo2018, Xu2019, Li2020, Yang2020}. In brief, a nanometer-thick layer of water near an interface possesses a modified molecular structure and dynamics, similar to that water exhibits in AC fields~\cite{Artemov2019}, with variations depending on channel form, type of surface, and thermodynamic conditions. Generally, confined water has an excess charge, shows enhanced mass transport~\cite{Feng2016, Li2020}, and exhibits quantum effects~\cite{Ceriotti2016}. The interfacial layer is also self-screened~\cite{Backus2021} which implies a clear border with bulk water and allows its isolation. Theory predicts such behavior of water at the interface with a broad class of solid surfaces~\cite{Secchi2016, Bjorneholm2016, Mamatkulov2017, Gonella2021, Artemov2021}. However, the changes in dielectric properties of interfacial water are most commonly reported on carbon-based materials~\cite{Holt2006, Cicero2008, Artemov2020b, Li2020, Kavokine2022}.

Unique properties of water-carbon interfaces owe to the specific atomic structure of the carbon~\cite{wang2021}. It appears in different forms, such as nanotubes, graphene, fullerenes, amorphous carbon, graphite, and diamond. Each form of carbon has an individual structure, morphology, and dielectric properties. Thus, they affect the structure and dynamics of water in contact with them in a different way. The electronically conductive forms of carbon, such as nanotubes or graphene sheets, have been shown to lower the viscosity of the interfacial water layer~\cite{Holt2006, Xie2018, Kavokine2022}. Being under electric potential, they also allow a formation of nm-thick electric double layer (EDL) at water-surface interface~\cite{Israelachvili1983}. Although the structure of EDL in pure water is still lacking clarity, this layer was shown to retain a significant amount of energy. For instance, the theoretical specific capacitance of EDL in water in contact with graphene is 550 F/g~\cite{KE2016}, which is twice as large as that of activated carbon~\cite{Jing2021} so that the corresponding energy density of aqueous supercapacitors based on graphene reaches 100 Wh/kg and is comparable with that of lithium-ion accumulators but additionally yields a high power density due to the smaller thickness of aqueous EDL.

On the other hand, studies show that dielectric carbon materials, such as diamond, increase protonic conductivity and polarizability of interfacial water~\cite{Batsanov2012, Fumagalli2018, Artemov2020a}. The contact of water with the dielectric form of carbon intensifies its electrodynamics in a thin nanometer layer only, but the ionic (protonic) conductivity of the interfacial water exceeds that of bulk water by five orders of magnitude~\cite{Artemov2020a}. In other words, the diamond surface (and probably other types of dielectric surfaces) makes interfacial water behave as a strong electrolyte~\cite{Artemov2021}, allowing a route for fast charge transfer along the interface so that protons and proton holes are moving in opposite directions. If one extends this route to the electrodes, effective charge separation can be achieved even with pure water as an electrolyte. The challenge is to find an optimal membrane-electrode assembly (hereinafter cell) to combine high charge transfer with the high storage capacity of interfacial water.

\begin{figure}
    \centering
    \includegraphics[width=\linewidth]{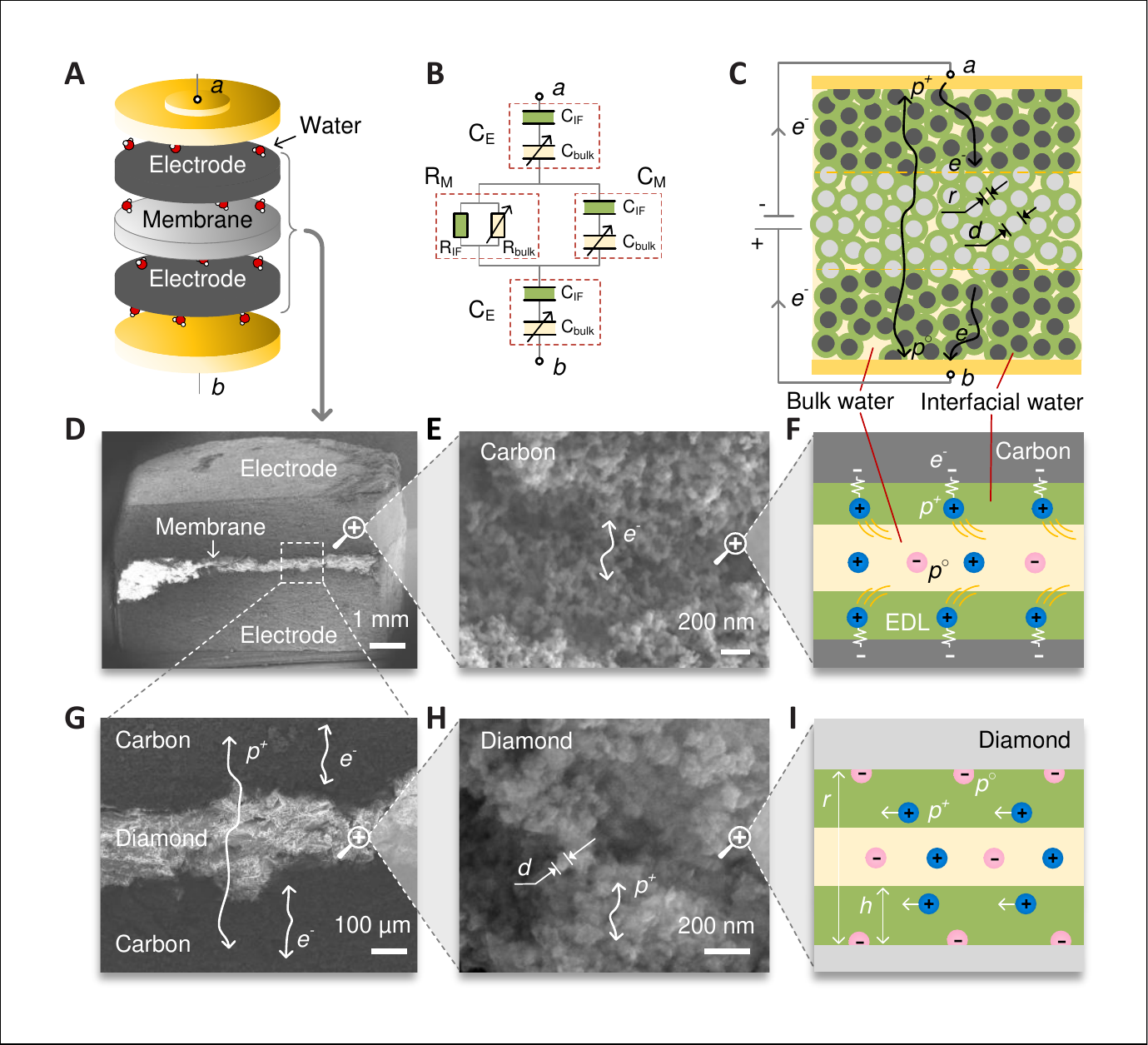}
    \caption{\textbf{Schematic of the sustainable confined-water accumulator.} (\textbf{A}) Layout of the carbon-based functional layers (cell). (\textbf{B}) Equivalent electric circuit of the cell. The letters $a$ and $b$ denote the points where the impedance measurements were applied. (\textbf{C}) Structural model of the cell. Circles show grains of carbon (dark gray) and diamond (light gray) of variable diameter \textit{d}. The pore space is filled with water. Interfacial water is shown in green and bulk water is shown in yellow. Arrows show the route for electronic and protonic conductivity. (\textbf{D}), (\textbf{E}), (\textbf{G}), and (\textbf{H}) Electronic microscopy micrographs of the functional layers at different resolutions (magnifications: 27$\times$, 25000$\times$, 120$\times$, and 125000$\times$, respectively). (\textbf{F}) Model of the interaction of water with the charged carbon surface. Plus and minus signs represent excess protons (H$_3$O$^+$ ions) and proton holes (OH$^-$ ions). (\textbf{I}) Model of the interaction of water with the diamond surface. Plus and minus signs represent the same species as mentioned above. The white arrows with \textit{r} and \textit{h} show the pore size and the interfacial water thickness, respectively.}
  \label{fig:1}
\end{figure}

\subsection{Confined-water double-layer capacitor}

Figure~\ref{fig:1}A shows a cell of carbon-based materials we fabricated for studying the charge storage capacity of nano-confined water. Each cell was made of two electrodes and a dielectric separator (Fig.~\ref{fig:1}, D and G). All layers were pure carbon materials: fine-grained activated carbon and nano-diamond, for the electrodes and the membrane, respectively (Fig.~\ref{fig:1}, E and H). We fabricated several dozens of such devices with six selected membrane grain sizes: 5, 40, 80, 120, 200, and 500 nm, varying the pore size of the separator layer over two orders of magnitude (see Fig.S1 in SI). All grains had narrow size distribution with a dispersion of less than 40\% and were compressed to a porous ceramics structure with the packing close to the densest one. Note that for the random-shape closely-packed particles, the mean pore size $r$ is about three times smaller than the mean grain size $d$, and the net pore volume is about 35\% of the total volume~\cite{Bernal1961}.

Unlike in previously used carbon-based membrane-electrode assemblies~\cite{Huettner2021, Krishnan2021}, our cell had several distinctions. The separator was made of nano-grained dielectric material and placed between the electrode layers with no gaps between layers. The grains of the electrode and the separator were tightly adjoined to each other, forming a continuous percolation network of nanopores without interruption of the thin interfacial water layers on the layers' boundaries (Fig.~\ref{fig:1}C). The pores were filled with water, which served as a medium for protonic conduction. The equivalent scheme of the cell (Fig.~\ref{fig:1}B) was represented by two capacitors $C_E$, and the capacitor-resistance pair, $C_M$-$R_M$, for electrodes and the membrane-separator, respectively. To calculate the parameters at different pore sizes, water confined to the nanopores was considered as that split into two parts (Fig.~\ref{fig:1}C): the interfacial water (green) and the bulk water (yellow). As the interfacial and the bulk water have different dielectric properties, each element of the equivalent circuit was represented as a combination of resistances, $R_{bulk}$ and $R_{IF}$, and capacitances $C_{bulk}$ and $C_{IF}$, representing the bulk and the interfacial water layers, respectively.

\subsection{Electrochemical performance of confined-water capacitor}

\begin{figure}
    \centering
    \includegraphics[width=\linewidth]{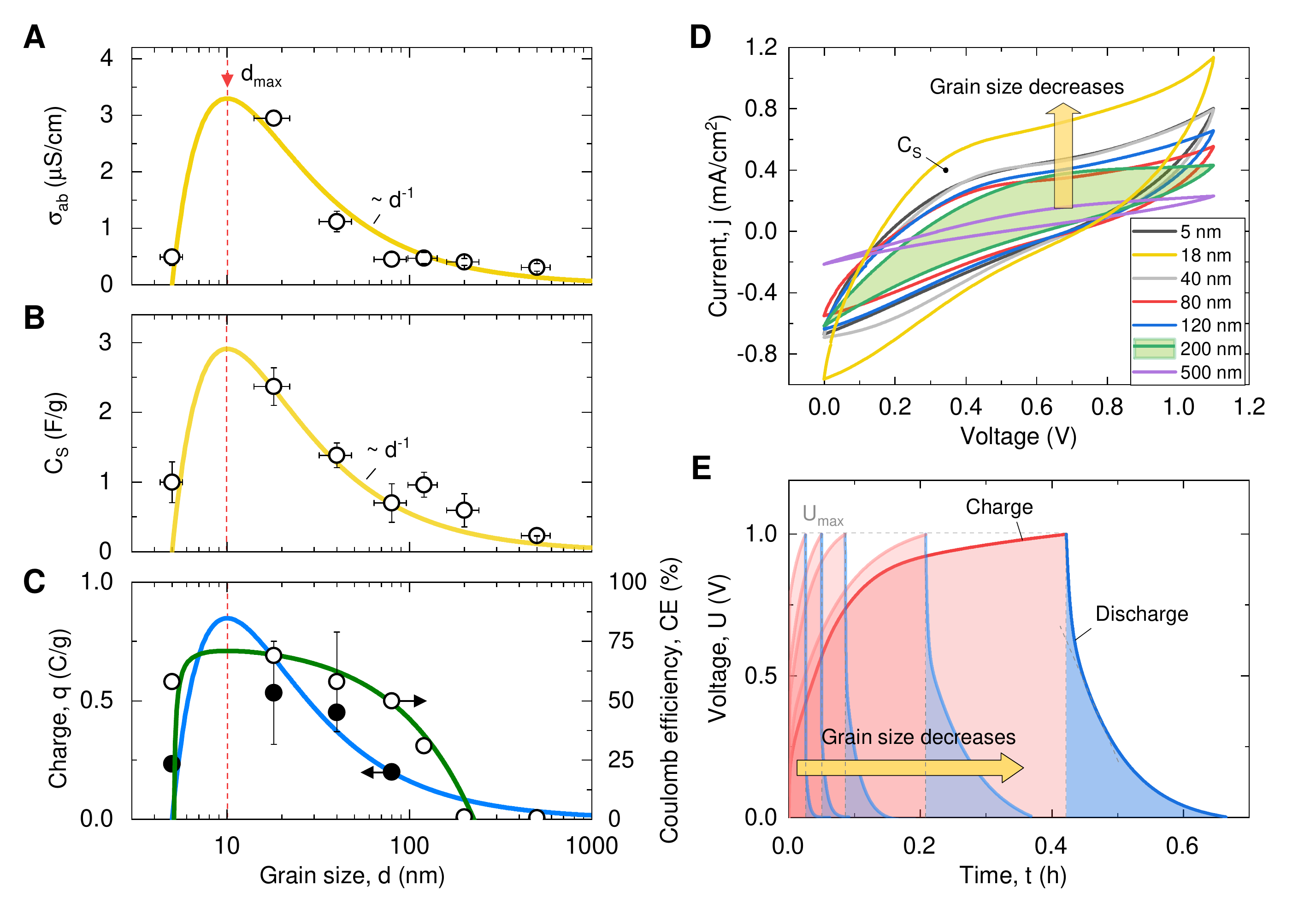}
    \caption{\textbf{Electrodynamic characteristics of sustainable confined-water accumulator.} (\textbf{A}) Protonic DC conductivity, $\sigma_{ab}$, of the cell, measured between points $a$ and $b$ (see Fig.~\ref{fig:1}B). The dots are experiments and the yellow lines are models according to the model discussed in the text. (\textbf{B}) Specific capacitance of the cell assembly calculated from the voltammograms. (\textbf{C}) Specific charge, $q$, and Coulombic efficiency, $CE$, of the cell. (\textbf{D}) Cyclic voltammograms of the cell assembly for different grain sizes (see legend). The area of the loops corresponds to the cell-specific capacitance C$_s$ (see panel \textbf{B}). (\textbf{E}) Voltage evolution on a charge-discharge cycle at a current j=0.4 mA for various grain sizes.}
\label{fig:2}
\end{figure}

Figure~\ref{fig:2} shows the dependence of electrodynamic characteristics of the cell on the grain size $d$: ionic (protonic) conductivity $\sigma_{ab}$ (Fig.~\ref{fig:2}A), capacitance $C_s$ (Fig.~\ref{fig:2}B), charge density $q$, and Coulomb efficiency $CE = (1-j\cdot \sigma_{ab}^{-1}U_{max}^{-1}) \times 100\% \approx t_{d}/t_{c}\times 100\%$, where $t_{c}$ and $t_{d}$ are the time of charge and discharge, respectively (Fig.~\ref{fig:2}C). All the characteristics change synchronously and are inversely proportional to $d$ for large grain (pore) sizes. This behavior is observed down to $d \approx$ 20 nm. For smaller $d$, the values reach a maximum and saturate. The maximum values of conductivity $\sigma_{ab}$ and capacitance $C_s$, which are related to Coulomb efficiency and the charge $q$, are expected for the gran size $d_{max} \approx$ 10 nm, or the pore size $r_{max} \approx$ 3 nm (see Discussion section). A further decrease of $d$ down to a few nanometers leads to a sharp drop in all the cell characteristics, presumably due to the overlap of the interfacial water layers from the opposite sides of the pore walls and a subsequent Coulomb blockade of the ionic charge carriers~\cite{Feng2016}. Note that the load of the pores was completely filled with water for all the grain sizes, including the smallest one, and was controlled by the weight comparison of the dry and wet cells.

The cyclic voltammograms of the cells (Fig.~\ref{fig:2}D) strongly depend on the grain size $d$. The areas of the hysteresis loops are proportional to the capacitance $C_s$ (Fig.~\ref{fig:2}B). For the large pores of fractions of microns, the loop opening is small due to the high resistivity caused by the low flow of ions (protons) through the separator. On the contrary, tiny pores have a more significant hysteresis due to the cumulative effect of a larger specific surface and an increase in the concentration of charge carriers passing through the membrane due to their unscreening near the interface~\cite{Artemov2020a}. The colored regions under the charge-discharge curves (Fig.~\ref{fig:2}E) also confirm an increase in capacitance following a decrease of $d$. The nonlinearity of the charge-discharge curves is associated with the equivalent distributed resistance (EDR) at the initial stage of the discharge process. EDR consists of two parts: equivalent series resistance (ESR) and additional resistance caused by charge redistribution in electrodes~\cite{Noori2019}. A small ESR here means that the resistivity of the interface between the carbon and diamond layers is small due to the interpenetration of water-filled channels between the electrodes and the membrane. The shape of the charge-discharge curves indicates that the charge is stored in an electrical double layer~\cite{Israelachvili1983}, which here is formed by separating excess protons and proton holes, or, H$_3$O$^+$ and OH$^-$ ions. Note that the way we filled pores with water excludes the addition of foreign contaminants. The conductivity changes due to the increase of the fraction of pure water, which structure is affected by the electrostatic interaction with the surface~\cite{Artemov2016, Artemov2020a}.

\section*{Discussion/Conclusion}

The behavior reported above can be understood using the model of the cumulative effect of bulk and interfacial water layers inside the interconnected network of nanopores. We consider the inner pore space as an effective medium formed by a constant-thickness interfacial water layer and the bulk water layer (Fig.~
\ref{fig:1}, F and I). When the pores are of hundreds of nanometers, the ratio between $r$ and $h$ is large enough for the interfacial water layer of about one nanometer to be neglected. In this case, the pore dielectric properties are determined by the bulk water properties. In small nanopores of a few nanometers, $h$ is comparable with $r$, and the cell parameters are determined by interfacial water instead. In a more general case of intermediate pore sizes, the model yields a common formula for the conductivity $\sigma_{ab}$ and capacitance $C_s$: $F$($r$)=$A$+$B h$($r$-$h$)$r^{-2}$, where $F$($r$) is either $C_s$($r$) or $\sigma_{ab}$($r$), $A$ is either $C_{bulk}$ or $R_{bulk}^{-1}$, and $B$ is either $C_{IF}$ or $R_{IF}^{-1}$ (see Fig.~\ref{fig:1}B and Fig.S7 in SI).

The parameter $h$ in the equation above, which determines the effective thickness of the interfacial water layer, is the only unknown argument. Previous studies reported thickness of the interfacial water layer $h \approx$ 1.5 nm~\cite{Israelachvili1983, Lee1984, Toney1994, Cicero2008, Tocci2014, Velasco2014,Fumagalli2018, Artemov2020a}. This value also coincides with the mean distance between the short-lived intrinsic ions of water~\cite{Artemov2020b, Artemov2021}. The best fit of the model to the experimental data with this value of $h$ is shown in yellow in Figs.~\ref{fig:2}, A and B. The parameters are given in Table S1 in SI. Note that although there are no data points near the curves' maxima, the simultaneous fit of $\sigma_{ab}$, $C_s$, $CE$, and $q$, unambiguously gives $r_{max}$ = 3 nm. This value is twice as large as $h$. Thus, the maximal conductivity and the capacitance are observed when all the pore space is filled with the interfacial water, but the layers from the opposite sides of the pore walls do not yet overlap. As such, nano-confined water can be treated as a superposition of bulk-like water and interfacial water, whose cumulative properties depend on their volume fractions. The size effect on the electrical conductivity of confined water is equivalent to the frequency increase\cite{Artemov2019} or to adding electrolytes\cite{Artemov2016}. However, the high conductivity and capacitance are reached here without using foreign species that positively distinguish our device over other electric energy accumulators.

\begin{figure}
    \centering
    \includegraphics[scale=0.4]{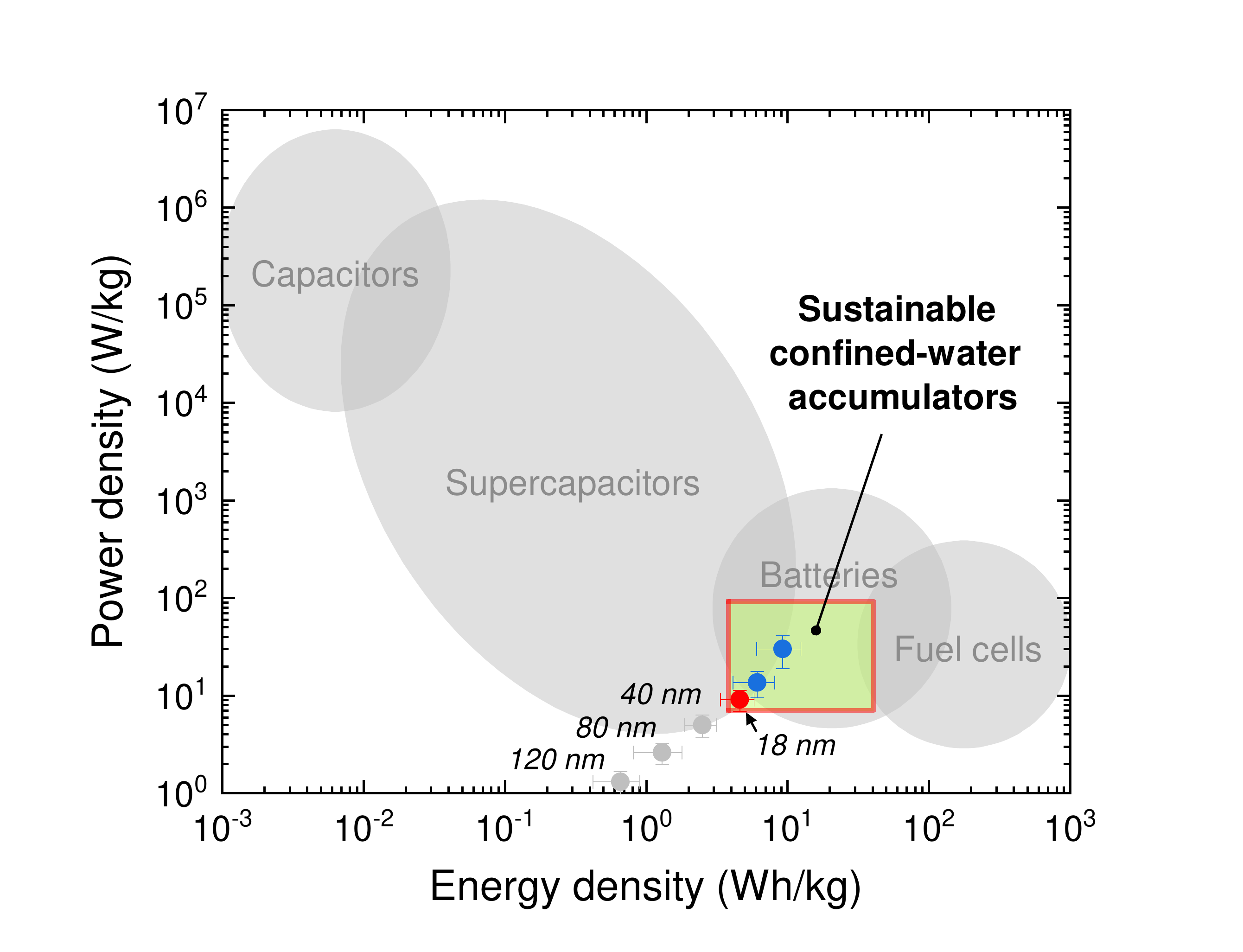}
    \caption{\textbf{Electric energy sources by power and energy density.} The dots show the experimental parameter obtained in this study for the cell with only pure water as an electrolyte. The gray points correspond to the cells with different pore sizes (see the numbers) of the thick model separator of 1 mm. The red point corresponds to the best parameters of the model cell with pore sizes close to ideal. The blue points are for the same cell but a thinner separator of 0.5 and 0.3 mm. The green rectangle shows the area the sustainable confined-water accumulators can occupy in case of optimization suggested in this research. The borders of gray regions are based on state-of-the-art experimental data.}
\label{fig:3}
\end{figure}

Figure~\ref{fig:3} compares the characteristics of our device with that known for different energy sources: accumulators, batteries, and fuel cells. Our water-based electrical energy accumulator allowed power and energy density of 5 W/kg and 2.5 Wh/kg, respectively (see the red dot). However, these values were obtained for the membrane thickness of 1 mm, which is good for the model experiments because it allows robust experimental conditions. Still, it is relatively thick for practical applications. We made several assumptions based on the known experimental data to estimate our water-only device's maximum possible energy and power density. First, the technologically plausible minimum thickness of the separator is 10 $\mu$m, typical for commercial supercapacitors. Second, the proton's mobility is up to 10 times higher than that of other ions due to the Grotthuss transport mechanism. The small size of the proton and the corresponding quantum effects~\cite{Ceriotti2016} also allow for further optimization beyond using foreign ionic species. Third, novel 2D materials suggest up to three times larger surface area per unit of mass compared to other carbon-based materials~\cite{Stoller2008} and offer up to 10 times higher density of the adsorption sites for the protons than for other ions~\cite{KE2016}. Considering the cumulative effect of these properties, the water-only system could exceed the specific energy range of commercial supercapacitors~\cite{Salanne2016}, becoming comparable with that of batteries, including metal-ion batteries~\cite{Tian2021}.

Thus, our water-based nano-structured electric energy accumulator demonstrates reasonable capacitance and, most importantly, combines them with neutrality to the environment. A new approach to the construction of aqueous energy systems discussed in this paper provides an excellent opportunity for sustainable devices to be improved and used in stationary and portable applications, although optimization is needed before commercialization. Further studies of the water interaction with topological 2D materials, including studying interfacial water as a 2D liquid, will shed additional light on the microscopic mechanisms of the anomalous properties of water at the nanoscale. We anticipate an increase in the effective surface area of the electrodes for up to an order of magnitude compared to the current carbon-black electrode materials, which leads to an additional increase in energy density. Recent studies suggest some promising directions regarding the optimization of the electrodes and functional materials~\cite{Krishnan2021, Chmiola2006}.

In conclusion, we have tested the charge storage ability of pure water confined to the nanopores of carbon-based materials by assembling the corresponding membrane-electrode block and found it promising for potential applications in ecologically-friendly energy-storage systems. We associated the ability of water confined to few-nanometer pores to transfer a significant amount of charge per period of time with the mechanism of descreening short-lived intrinsic ions of water near the interfaces in a layer of about 1.5 nm (interfacial water). This effect was observed near carbon-based materials, but the mechanism seems material-independent. It is reasonable to expect similar behavior of water in the broader class of porous systems, including nanoporous clays and polymers, which can reduce the cost of energy storage. Although the anomalous properties of water in confinement/interface have been used here for energy storage applications, our understanding of interfacial phenomena could potentially shed light on biological phenomena such as intercellular transport, neuroactivity, or molecular translocations.

\section{Methods}

To probe the charge storage ability of the cell, we used dielectric spectroscopy, cyclic voltammetry, and galvanostatic techniques. In short, by applying an AC electric field, we tested the electric impedance and obtained the electric conductivity and the dielectric constant, connected with the mobility of the protons and the polarizability of the water-solid interface. Electrochemical measurements were conducted using a commercial potentiostat (see SI for details). To ensure the purity of water in the pores, they were filled by capillary condensation in the atmosphere of saturated water vapor. The surfaces of carbon and diamond nanoparticles are cleaned by centrifugation of nano-powders in distilled water multiple times, followed by drying at a high temperature (see SI for details). The integrity of the cell has been realized by mixing 5-weight \%  PTFE into the diamond powder (Fig.~\ref{fig:1}D). Note that the fraction of water confined to the pores was approximately the same for all fabricated cells independently of the grain sizes $d$ due to the close packing, while the size of percolating nano-volumes of water was changing accordingly.

\section*{Author Contributions}
A.Ry. and V.G.A conceived the idea of the experiment and wrote the manuscript. A.Ry. and S.M. built the experimental setup and conducted the electrochemical experiments. S.M., A.K., and T.W. prepared and characterized the samples. A.Ra and V.G.A. suggested a conceptual interpretation of the results. A.Ry, S.M., A.K., and A.Ra. analyzed the data with contributions from K.J.S. and V.G.A. K.J.S. and V.G.A. supervised the study. All the authors discussed the results and contributed to the final version of the manuscript.

\begin{acknowledgement}
We thank Pavel Kapralov for his interest in the topic and technical support, Federico Ibanez for fruitful discussions, Natalia Gvozdik for assistance with potentiostat measurements, Mariam Pogosova for useful advice, and Andrey Chernev for the help with TEM measurements. This work was supported by the Skoltech-MIT NGP project and the Horizon 2020 CompBat project of the European Commission.

\end{acknowledgement}

\section{Data availability}
All data are available within the paper and Supplementary Information. Source data are provided in this paper. The Supplementary Information includes Samples preparation and treatment, Measurement methods, Figures S1 to S10, and Table S1.

\bibliography{references}
\end{document}


\textbf{This PDF file contains:}

1. Samples preparation

2. Samples cleaning

3. Surface area determination

4. Filling cells with water

5. Electrochemical measurements

•	Figures S1 – \textcolor{black}{S10}

•	Table S1

\subsection{1. Samples preparation}

\begin{figure}
    \centering
    \includegraphics[scale=0.5]{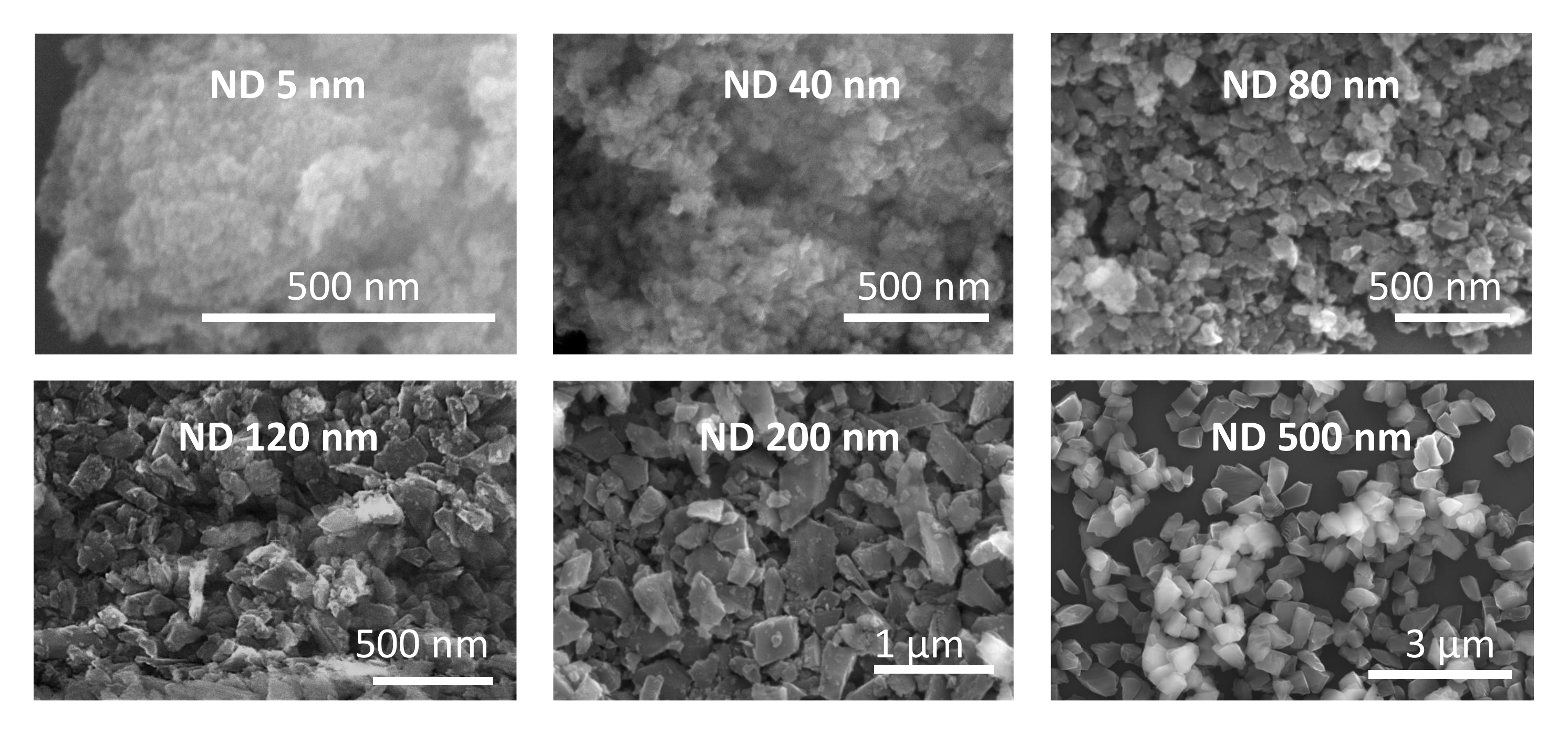}
    \caption{\textbf{Scanning Electron Microscope (SEM) images of nanodiamond (ND) powders of different grain sizes.}}
\label{fig:SI_1}
\end{figure}

\begin{figure}
    \centering
    \includegraphics[scale=0.5]{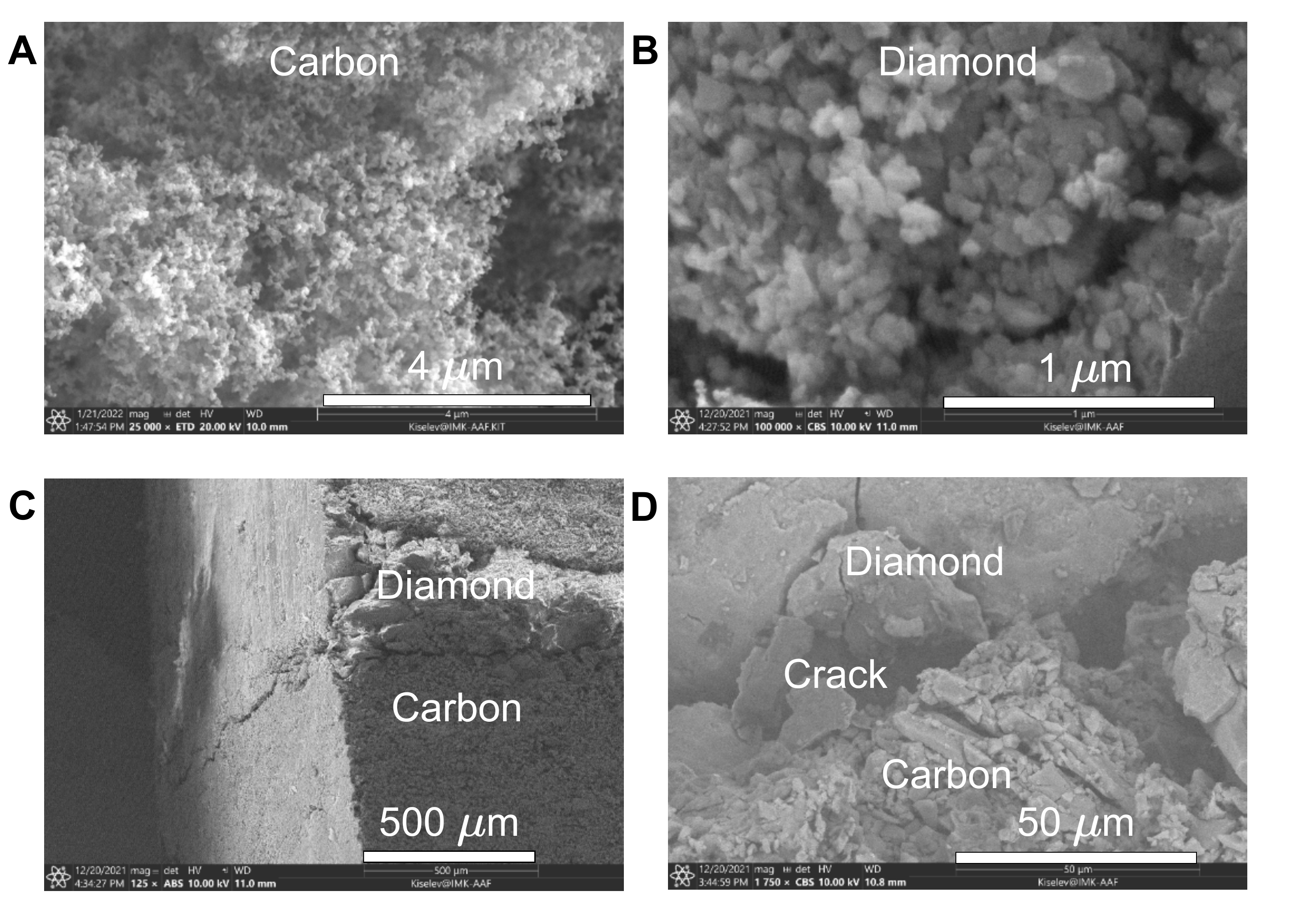}
    \caption{\textbf{Scanning Electron Microscope (SEM) images of the cell.} (\textbf{A}) Carbon electrodes. (\textbf{B}) Diamond membrane (80 nm). (\textbf{C}, \textbf{D})  Different cross-sections of the cell assembly.}
\label{fig:SI_2}
\end{figure}

Nanodiamonds powders with narrow grain size distributions around 5, 18, 40, 80, 120, 200, and 500 nm were purchased from Adámas Nanotechnologies (Fig. S\ref{fig:SI_1}). Carbon black powder with grain-size distribution around 40 nm was purchased from Sigma-Aldrich. The powders were carefully cleaned before use (see next section). Electrochemical heterostructures (cells) were pressed from the powders in a stainless-steel press form and consisted of 3 layers (electrode-membrane-electrode) approximately 1 mm thick each and 5 mm in diameter. We used a standard press form and a pressure of up to 20 tons per cm$^2$. Figure S\ref{fig:SI_2} shows the structure of the cell and layers at different scales. To increase the mechanical stability of cells, carbon particles were mixed with 5\% wt. of polytetrafluoroethylene (PTFE) or polyethylene glycol (PEG). Note that no difference was found between these 'glue' materials, indicating a lack of effect on confined water dielectric properties. A minimum of 15 cells of each grain size (about 100 cells in total) were fabricated to verify reproducibility and minimize the error bars of data points. Cracks (Fig. S\ref{fig:SI_2} C, D), and short-circuit between the electrodes, were observed for some samples, leading to a decrease in cells performance or a partial loss of percolation, or enlarged effective resistance of the cell. These samples were excluded from the data analysis.

\subsection{2. Samples cleaning}

The powders of nanodiamonds and carbon black were preliminary cleaned multiple times to avoid the influence of surface contaminants on the dielectric properties of water confined to the matrix of the pores formed by the grains of the host materials. To do this, the obtained from the manufacturer powders were dissolved in deionized water. We used Milli-Q Direct Water Purification System and 18 MOhm$\cdot$cm water. Then the mixture was centrifuged to separate the water and the powders, and the pH and DC conductivity of the brine were measured. The procedure was repeated many times until a complete stabilization of the measurements. Typically, about 10 cycles were needed to obtain no further change of conductivity and pH value within the accuracy of 0.05 units. After the washing, the samples were annealed in a laboratory oven at 200$^o$C. The analysis of washed powders consisted of elemental analysis, infrared spectroscopy, and titration previously described elsewhere (see SI of the paper~\cite{Artemov2020a}). These methods revealed atoms of oxygen, silicon, and chloride on the surface of our carbon-based materials, but their concentration was less than 0.01\% of the concentration of native carbon atoms. Moreover, the normalized-on-the-surface-area concentration of the IR-active surface groups was shown to decrease with the decrease of the grain size $d$, while the ionic (protonic) conductivity of water between the grains of these materials was shown to increase instead. Thus, the dielectric properties of cells were attributed to the confinement-changed properties of water, but not to the surface chemistry.

\subsection{3. Surface area determination}

\begin{figure}
    \centering
    \includegraphics[scale=0.5]{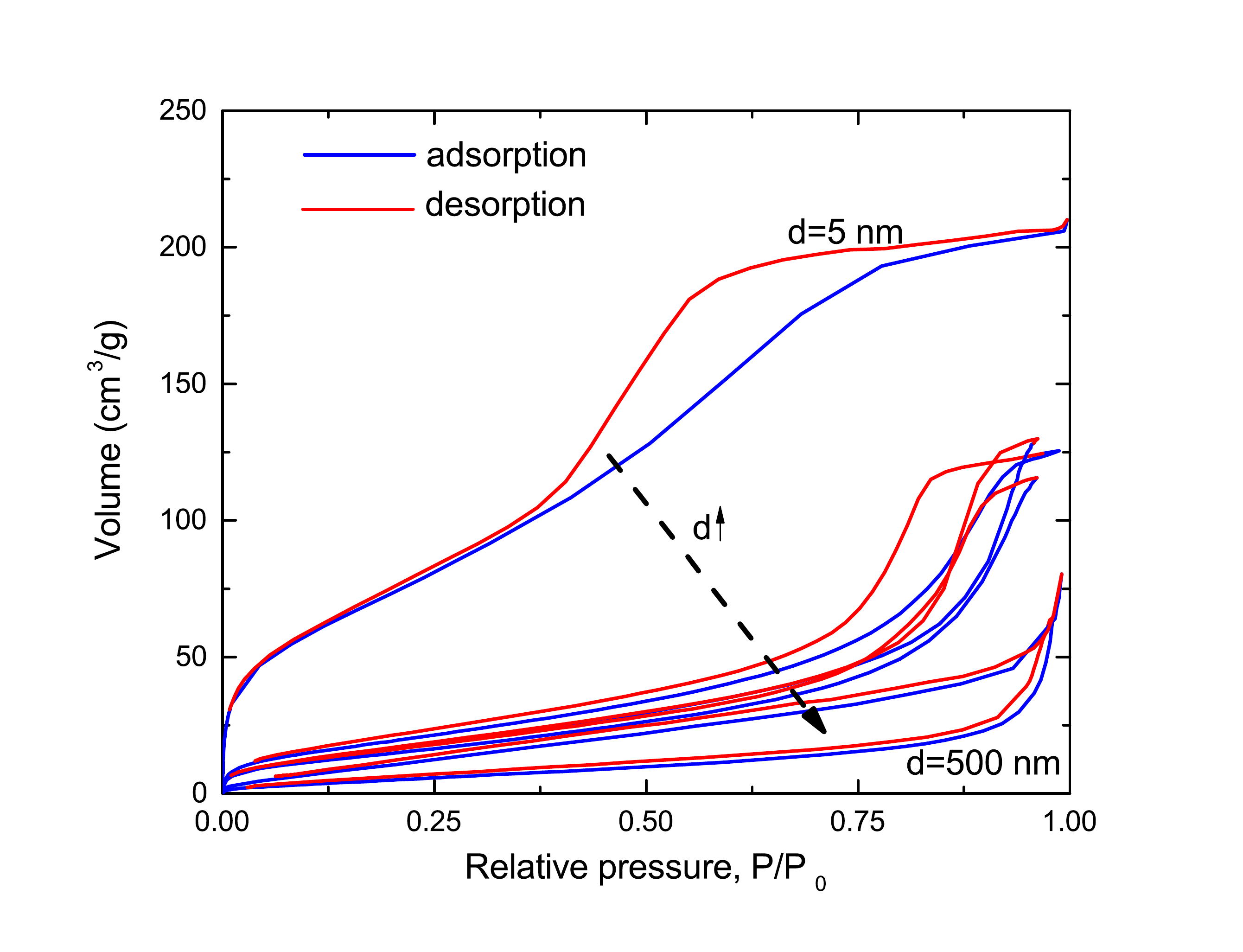}
    \caption{\textbf{Adsorption/desorption isotherms of argon on nanodiamond membranes.} Samples exhibit mesoporosity as indicated by the adsorption/desorption hysteresis. A sharp increase in the volume of adsorbed gas at $P/P_0>0.9$ indicates a capillary condensation.}
\label{fig:SI_3}
\end{figure}

\begin{figure}
    \centering
    \includegraphics[scale=0.5]{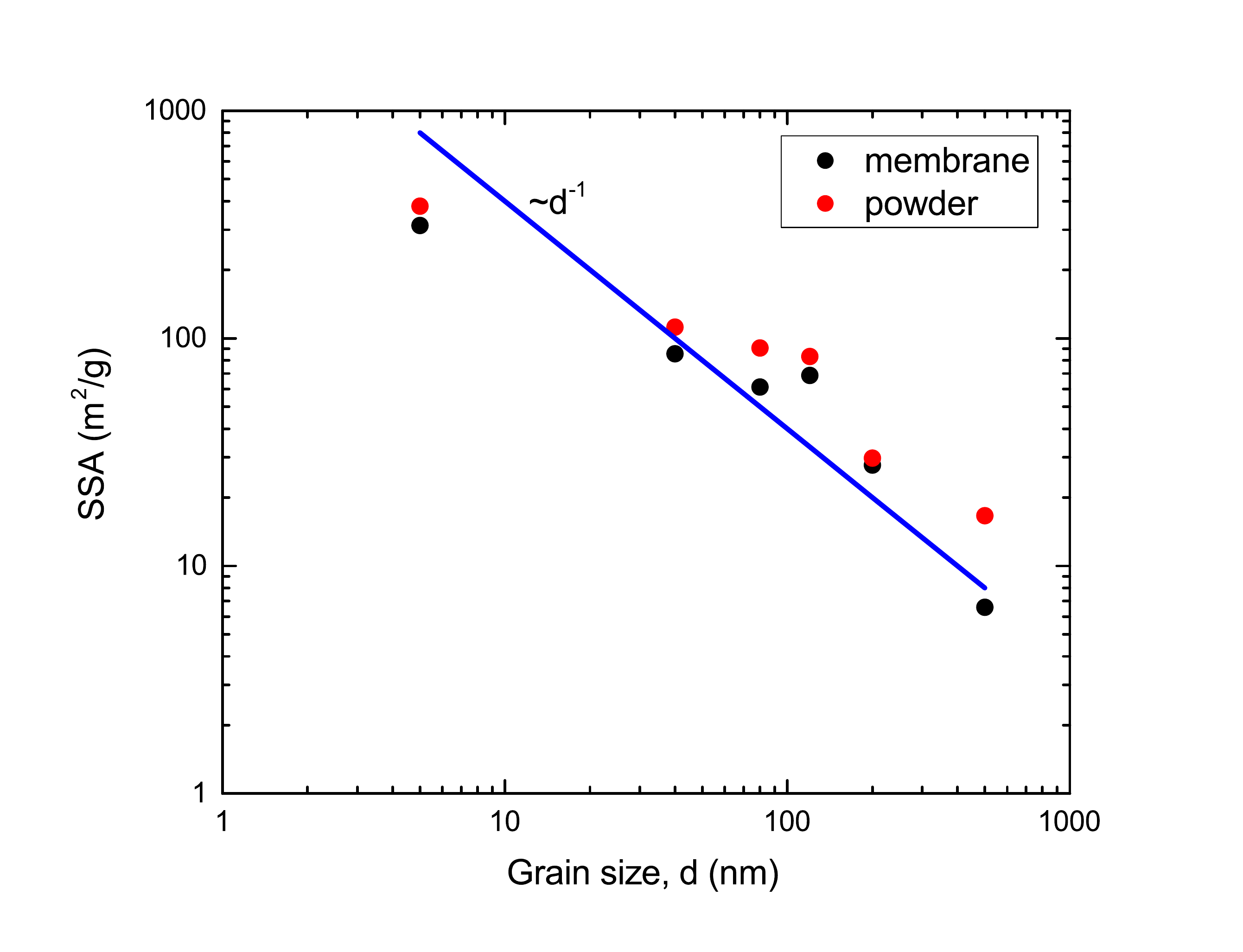}
    \caption{\textbf{Specific surface area (SSA) of nanodiamonds.} The red dots are for powder, and the black dots are for pellets made of powder. The SSA was calculated from Ar adsorption isotherms at 87.13 K (see Fig.~S\ref{fig:SI_3}). The blue line corresponds to the surface-to-volume ratio ($d^{-1}$) of the grains typical for spherical particles.}
\label{fig:SI_4}
\end{figure}

The Quantachrome Autosorb instrument was used to determine the specific surface area (SSA) of the samples. The argon gas adsorption/desorption isotherms were measured at 87.13 K (Fig. S\ref{fig:SI_3}). The SSA was calculated using standard Brunauer–Emmett–Teller (BET) analysis and was measured for both the powders and the compressed ceramics (pellets or membranes) as a function of the grain size $d$ (Fig. S\ref{fig:SI_4}). An inverse proportionality of the SSA (see the blue line) on $d$ corresponds to an increase of the surface area of the samples proportionally to the rise of the surface area of the grains following the decrease of their diameter $d$. A good coincidence of SSA of powders and pellets was observed. Thus, the porosity of the pellets was open, and all the pores were available for the molecular species.

\subsection{4. Filling cells with water}

The porosity of our electrochemical cells was obtained by the comparison of the volume, mass, and density of the raw materials and the pressed heterostructures. It was between 35 and 40\%, which is close to the closely packed limit. The free space between the grains was filled with water and was controlled by gravimetric analysis. The following procedure was developed to assure a complete pore filling. The cells were placed in a desiccator and pumped within 30 minutes to remove the air from the pores. Then, samples were stored in the same desiccator in the atmosphere of the saturated water vapor for 10 hours. The mass gain was controlled gravimetrically and showed a complete occupation of the pores by water. The closed Teflon casing with inserted parallel-plate gold electrodes was used to prevent cell drying during the measurement. The cells' mass was controlled before and after the measurements and showed no decrease within several hours.

\subsection{5. Electrochemical measurements}

\begin{figure}
    \centering
    \includegraphics[scale=0.5]{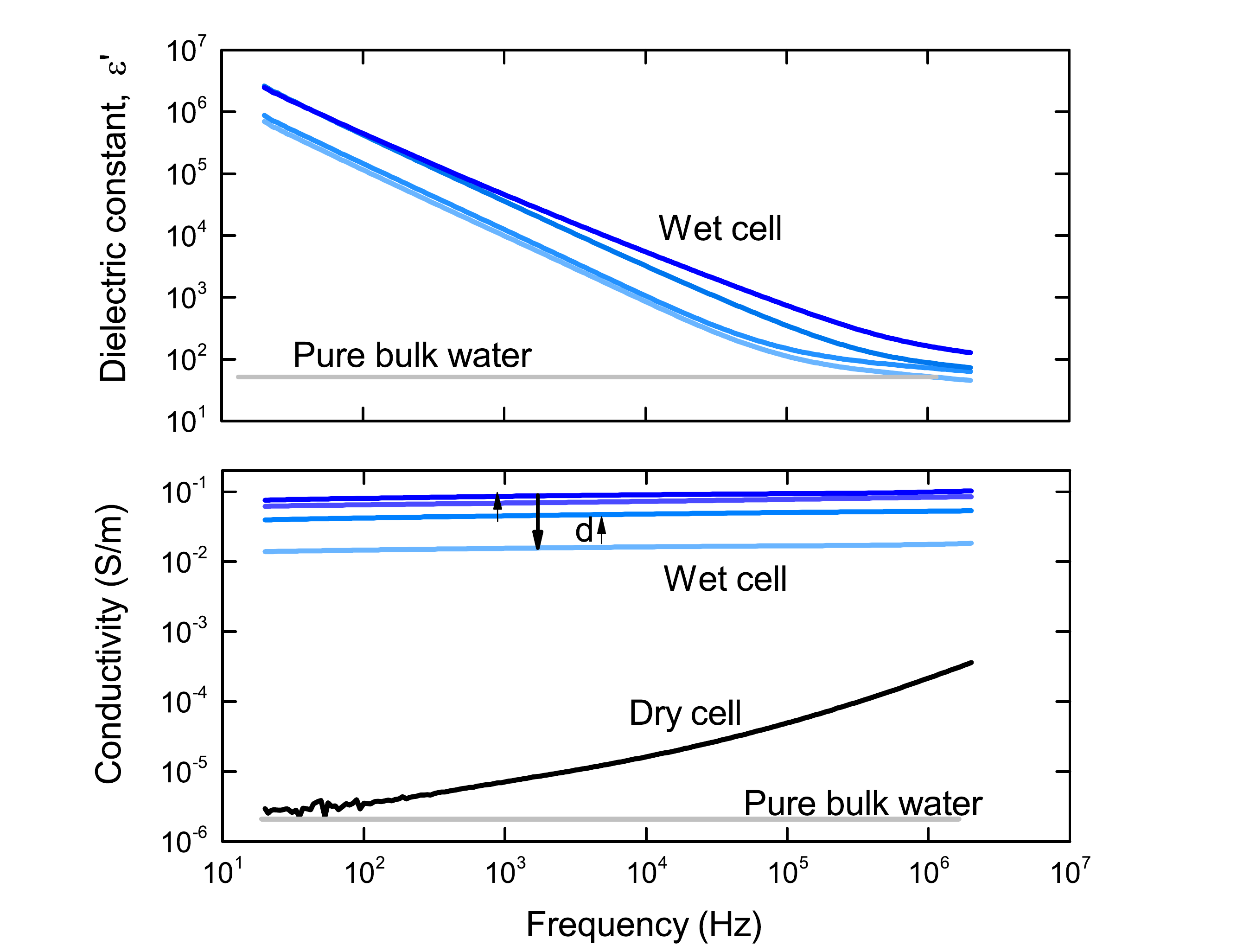}
    \caption{\textbf{Dielectric spectra of the cells in the frequency range from 20 Hz to 2 MHz.} The upper panel is for the real part of the dielectric functions, and the bottom panel is for the dynamic conductivity. The shades of blue correspond to the water-filled porous cells with different grain sizes $d$. The black curve is for the dry cell. The gray horizontal lines are for the dielectric constant and the dynamic conductivity of pure bulk water.}
\label{fig:SI_5}
\end{figure}

The complex impedance (EIS) of the cells was measured using a Keysight E4980A analyzer operating in the frequency range from 20 Hz to 2 MHz. The parallel-plate-capacitor geometry with circular polished electrodes covered with gold was used. The electrodes' diameter was 5 mm, and the thickness of the cell was approximately 3 mm. All the measurements were performed at 22 $\pm$ 2 $^{\circ}$C (room temperature). Figure S\ref{fig:SI_5} shows the typical EIS data. The DC (see the low panel and low frequencies around 10 Hz) electric conductivity of the dry cell was approximately five orders of magnitude lower than that for wet samples. The electric conductivity of pure water is shown in gray for comparison and corresponds to 5.5 $\cdot$ 10$^{-6}$ S/m. The dielectric constant (see the upper panel and low frequencies), on the contrary, was up to five orders of magnitude than that of bulk water and up to seven orders of magnitude larger than that of dry sample (out of the graph, approximately 3 units of the dielectric constant). This additional polarization of wet samples is due to the diamond/water interface, presumably due to the formation of an electrochemical double-layer of intrinsic ionic species of water.

A BioLogic impedance analyzer was used for Cyclic Voltammetry (CV) measurements. Two-electrode configuration was applied to the same samples used in EIS experiments. The current density of the galvanostatic test was ramped from 1 to 20 mA/cm$^2$. The cell voltage was always maintained below 1 V to avoid massive hydrolysis. Each sample was subjected to several 0-1-0 V cycles with a scanning rate of 5 mV/s (see Fig. 2d of the main text). Voltammograms were used to calculate the capacitance, which is proportional to the hysteresis area. The latter was normalized by the voltage window. To obtain specific capacitance, the values were divided by the cell mass (Fig. 2c of the main text), i.e., Teflon casing and metal electrodes were excluded. 

\begin{figure}
    \centering
    \includegraphics[scale=0.5]{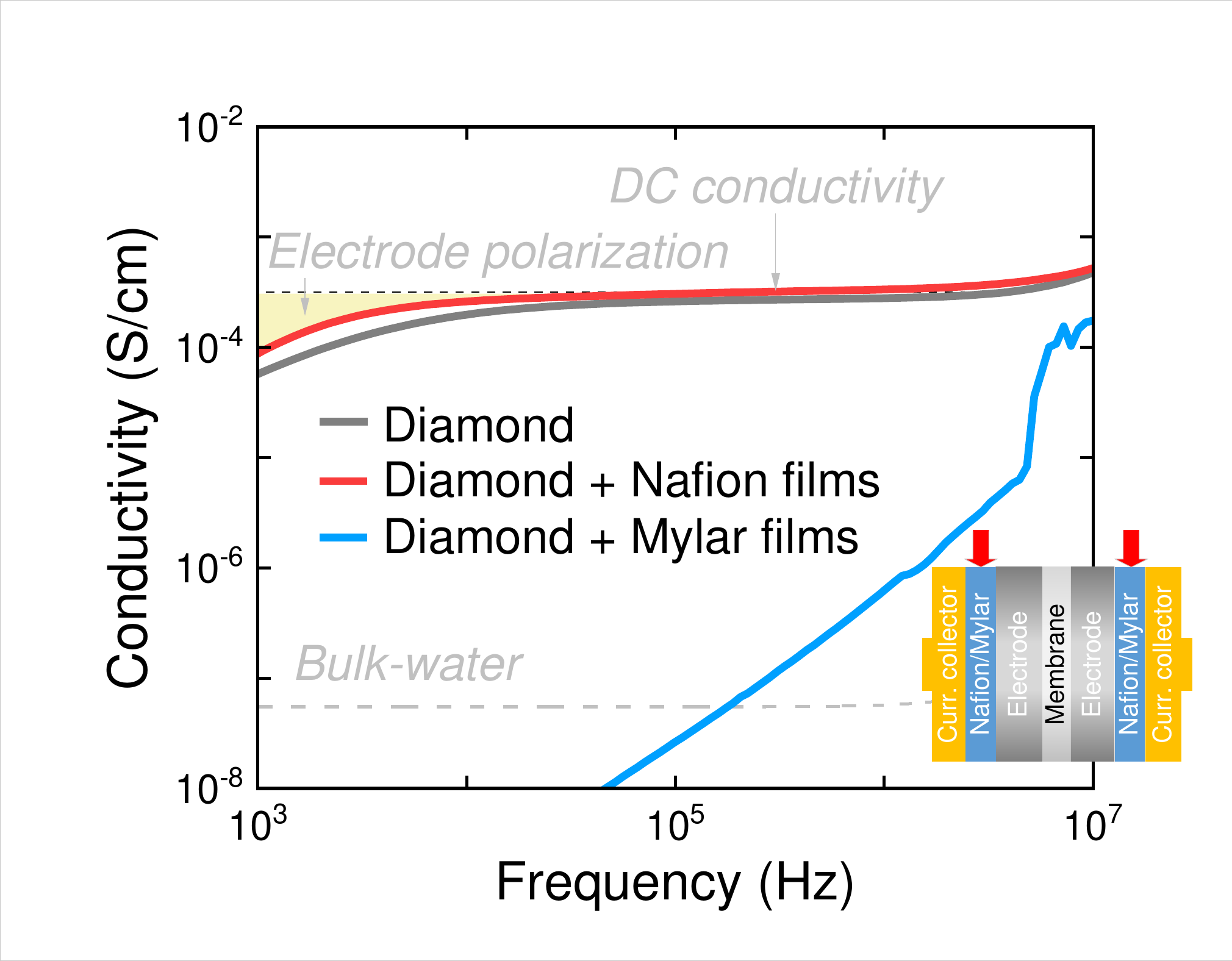}
    \caption{\textcolor{black}{\textbf{Electrical conductivity spectra}  (1 kHz to 10 MHz) of cells without blocking electrodes (grey), with Nafion (red), and with Mylar (blue) films placed between the cell and the electrodes (see inset). The conductivity of pure bulk water is shown by a dashed line for comparison. The inset scheme shows the layout of the experimental cell. Red arrows indicate the position of additional layers, which were placed to confirm the protonic conductivity.}}
\label{fig:SI_6}
\end{figure}

\begin{figure}
    \centering
    \includegraphics[scale=0.7]{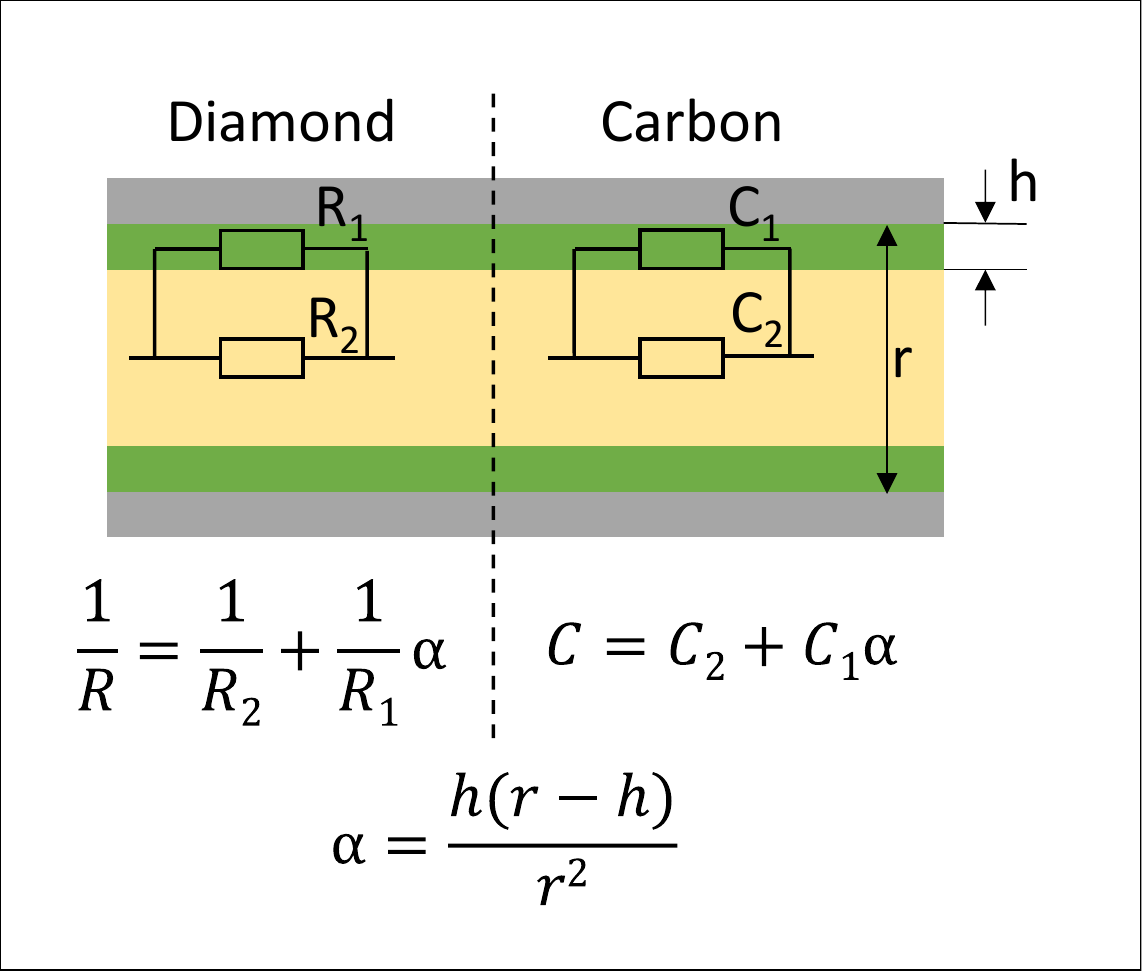}
    \caption{\textcolor{black}{\textbf{On the derivation of the model for conductivity and capacitance}  (see yellow lines in Fig.2, A and B, of the main text and the corresponding formula). Gray layers represent the walls of the diamond (left) and the carbon (right). Green and yellow layers are interfacial and bulk water, respectively. Formulas are for the equivalent resistivity (R) and capacitance (C), represented by the equivalent schemes. The coefficient $\alpha$ is a geometrical factor reflecting the pore's relative amount of bulk and interfacial water for a general case of the random pore shape.}}
\label{fig:SI_7}
\end{figure}

\begin{figure}
    \centering
    \includegraphics[scale=0.32]{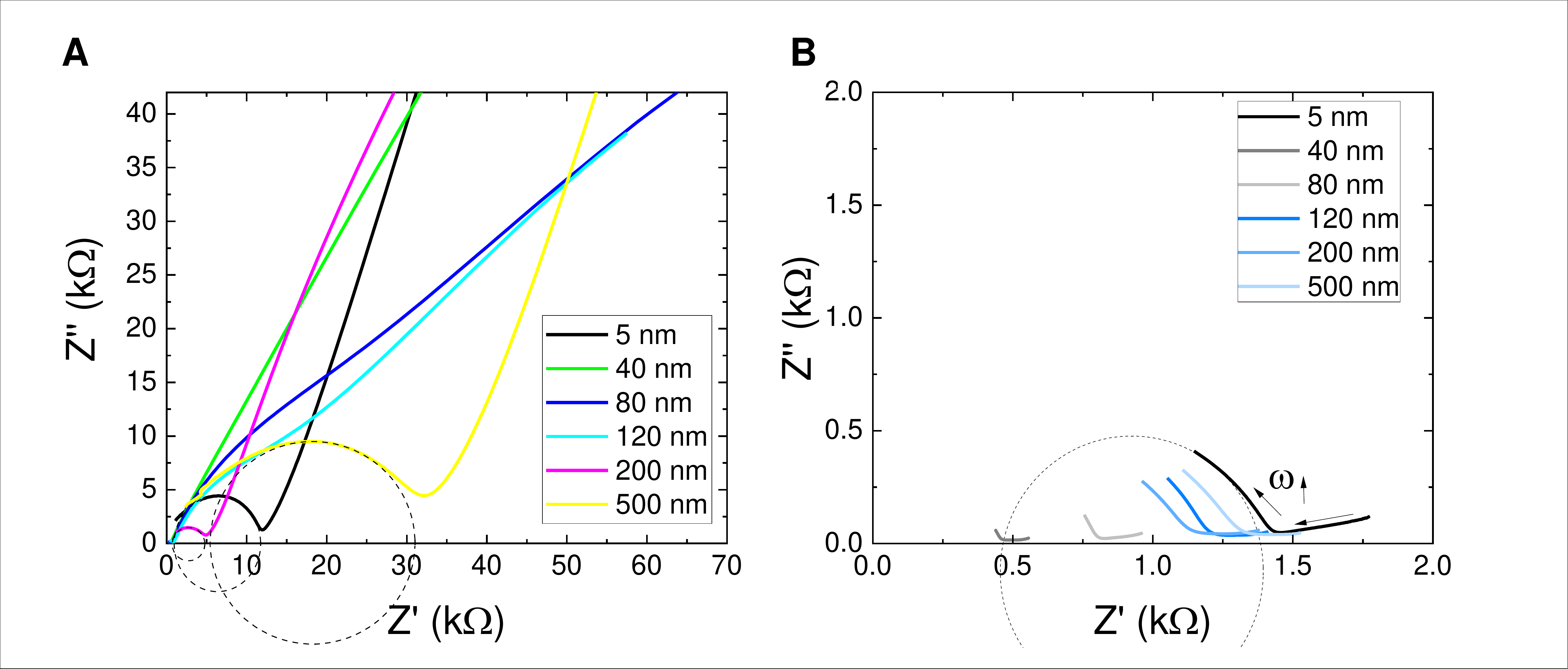}
    \caption{\textcolor{black}{\textbf{Nyquist plots.}  Complex impedance of the nanodiamond separator (A) and the carbon-diamond-carbon cell assembly (B). Different colors correspond to the different grain sizes of diamonds (see legend).}}
\label{fig:SI_8}
\end{figure}

\begin{figure}
    \centering
    \includegraphics[scale=0.32]{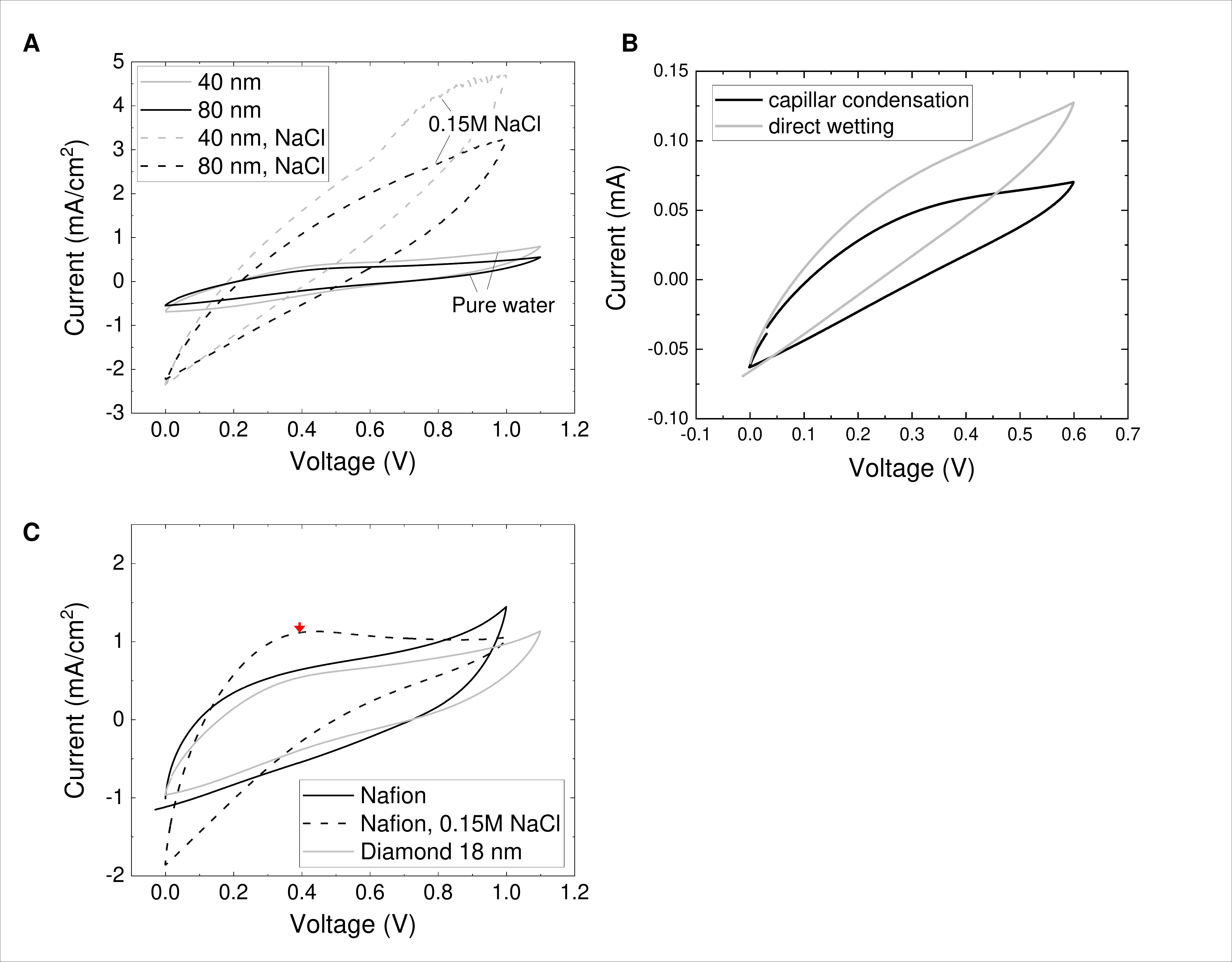}
    \caption{\textcolor{black}{\textbf{Cyclic voltammograms.} (A) Cell with a nano-diamond membrane filled with pure water (solid) and with a 0.15 M solution of sodium chloride (dashed). The gray and black curves correspond to 40 and 80 nm grain sizes. (B) Cell with a nano-diamond membrane filled by capillary condensation (black line) and direct wetting with tap water (gray). (C) Cell with a 1 mm nano-diamond membrane filled with pure water (gray), for supercapacitor with a 250-micron Nafion membrane filled with pure water (black solid), and with 0.15 M solution of NaCl (black dashed). The red arrow shows the part caused either by the stuck of the nonprotonic ionic species at the Nafion surface or by the parasitic redox reaction on the electrodes involving Na$^+$ and Cl$^-$ ions.}}
\label{fig:SI_9}
\end{figure}

\begin{figure}
    \centering
    \includegraphics[scale=1.0]{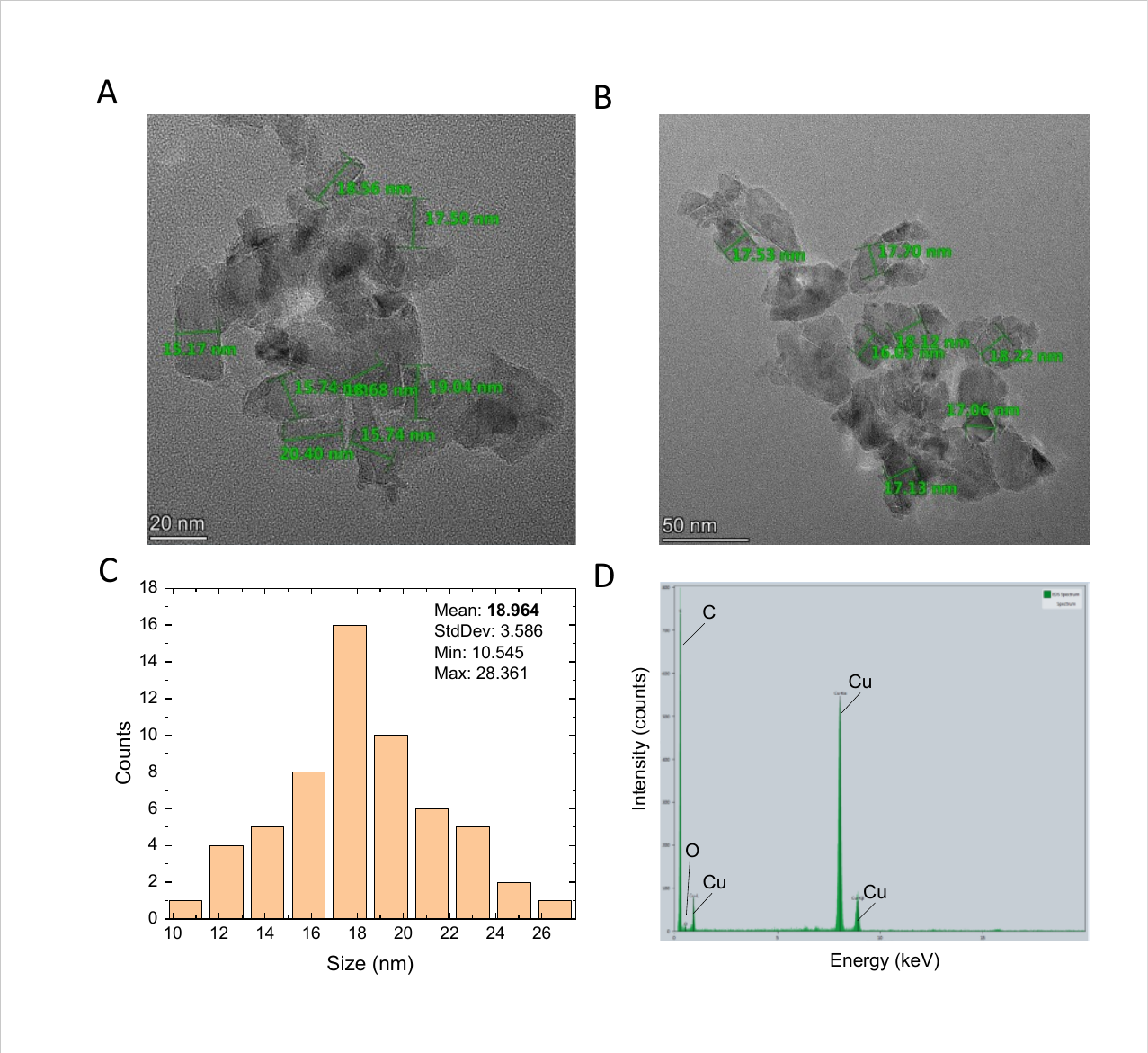}
    \caption{\textcolor{black}{\textbf{Characterization of 18-nm nanodiamonds}. (A and B) TEM pictures, (C) grain size distribution, and (D) EDS spectrum. Peaks of Cu on the spectrum correspond to the used substrate.}}
\label{fig:SI_10}
\end{figure}

\begin{table}[]
\caption{\label{tabS1}\textbf{Fit parameters of the model to the experimental data points in Fig.2.(see main text)}. $A$ is either $C_{bulk}$ or $R_{bulk}^{-1}$, and $B$ is either $C_{IF}$ or $R_{IF}^{-1}$ (see Fig. 1B of the main text)}
\begin{tabular}{lllll}
\cline{1-3}
\multicolumn{1}{|l|}{}  & \multicolumn{1}{l|}{Counductivity (S/cm)} & \multicolumn{1}{l|}{Capacitance (F/g)}     &  &  \\ \cline{1-3}
\multicolumn{1}{|l|}{A} & \multicolumn{1}{l|}{5.5$\cdot$10$^{-8}$}  & \multicolumn{1}{l|}{3.4$\cdot$10$^{-11}$} &  &  \\ \cline{1-3}
\multicolumn{1}{|l|}{B} & \multicolumn{1}{l|}{\textcolor{black}{1.5$\cdot$10$^{-2}$}}  & \multicolumn{1}{l|}{\textcolor{black}{11.5}}                 &  &  \\ \cline{1-3}
                        &                                           &                                           &  & 
\end{tabular}
\end{table}

\bibliography{references}